\documentclass[twocolumn,showpacs,preprintnumbers,amsmath,amssymb]{revtex4}
\usepackage{amssymb}
\usepackage[dvips]{graphicx}
\begin{document}
\noindent
\textbf{Heating-free I-V of the `intrinsic' Josephson effect}

Intrinsic tunnelling (ITs) is believed to occur in highly anisotropic superconductors such as Bi2212 and Bi2201, manifesting itself in the peculiar shape of an out-of-plane I-V. The low-bias brush-like part of such I-V is highly hysteresial and becomes single-valued at higher voltages, where gap-like peculiarities are located. Some important conclusions have been drawn from results obtained with this method cf.\cite{gough}, assuming Josephson coupling between Cu-O planes to be the {\it only} cause of the peculiar shape of I-V. This assumption, however, is not beyond dispute from the experimental viewpoint \cite{heat-1st,comment2201}. In particular, unlike in conventional tunnelling, the characteristic features of ITs are seen at high levels of dissipation (estimated as VI/A, where A is the area of the sample) in excess of 1kW/cm$^2$ at $\sim$4.3K \cite{comment2201}. This value significantly surpasses the critical heat load for liquid $^4$He, $\sim$1W/cm$^2$, thus signalling possible heating, especially since the poor thermal conductivity of all HTSC makes them particularly prone to local overheating. Indeed, significant heating was reported even for the `brush'-like part of I-V \cite{gough,heating}. Moreover, I have shown that the `intrinsic tunnelling' (IT) characteristics of Bi2212 and Bi2201 can be reproduced if the heating caused by Joule dissipation is the {\it only} reason for nonlinearities observed in their current-voltage characteristics \cite{heating,comment2201,crete-ije}. Thus, experimental  normal state c-axis resistance $R_N(T)$ and Newton's Law of Cooling provide a natural explanation of the divergency between the results obtained with conventional spectroscopy and those derived from IT.

Recently the conclusions of \cite{comment2201} were addressed by Yurgens in a way similar to \cite{gough}. Moreover, A. Yurgens \cite{reply} managed to reaffirm some of the experimental results of predecessors \cite{gough,heating}. Conservative estimates of overheating  \cite{reply} confirmed those  of \cite{comment2201} {\it quantitatively}. Unfortunately, instead of addressing the intrinsic (heating-free) shape of I-V and following \cite{gough}, Yurgens constructed a `heating-free' I-V using a variety of overheated I-V-T \cite{reply,aux}. In my opinion, these should be treated with caution until all sample-contact nonlinearities and overheating dependent temperature lags are addressed properly. This is especially important for cases of intermediate heating such as the `intrinsic Josephson effect'.

I will show that  the overheating at biases corresponding to hysteresial multi-branch I-V characteristics presented in Fig.8 from \cite{aux} allows for an estimate the intrinsic heating-free shape of I-V in the `intrinsic' Josephson effect. The original data are reproduced in the insert to Fig.1. Here, the last branch of I-V is plotted together with the temperature estimate, $T_s$, of the topmost electrode of this Bi2212 mesa-structure. T$_s$ increase is caused by the Joule dissipation inside the mesa \cite{comment2201}. The measured data provide little information on the `heating-free' I-V, addressed in the main panel of Fig.1, where the resistance of the sample is plotted together with its overheating above the base temperature, $T_0$. Fig.1 suggests the Ohmic response (shown by the dashed lines) to occur in the absence of overheating. At higher V$\geq$30mV we see a progressive rise of the overheating, which entails a corresponding reduction of $R_c$. This allows for an estimate of the highest heat load that this particular sample can accept at 4.3K without overheating, $W/A\leq0.4W/cm^2$ (A=6x6$\mu m^2$ is the area of the mesa). This is consistent with the  critical heat load for liquid $^4$He, $\sim$1W/cm$^2$. In my opinion the dotted lines in Fig.1 represent the intrinsic `heating-free' I(V), which has nothing in common with the Josephson behaviour.

\begin{figure}
\begin{center}
\includegraphics[angle=-0,width=0.47\textwidth]{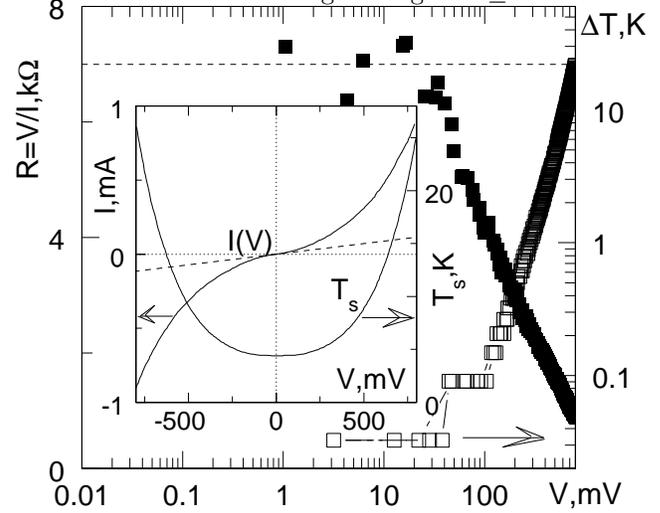}
\vskip -0.5mm
\caption{Mutually correlated variation of sample' resistance estimated for the return branch of I(V) in B-2212 mesa, Fig.8 from\cite{aux} (reproduced in the insert)  and its overheating, $\Delta T=T_s-T_0$, caused by the Joule dissipation, are shown by $\blacksquare$ and $\square$ respectively; dotted lines  represent a `heating-free' I(V). }
\end{center}
\end{figure} 

To conclude, novel results by Yurgens \cite{reply,aux} allow for an estimate of the intrinsic heating-free shape of I(V) in the `intrinsic' Josephson effect and offer additional support to the original conclusions of \cite{heat-1st,comment2201,crete-ije}. 

The financial support of the Leverhulme Trust (F/00261/H) is gratefully acknowledged.

V.N.Zavaritsky 

Department of Physics, Loughborough University, Loughborough LE11 3TU, United Kingdom.

\end{document}